\documentclass[a4paper,
               biblatex,     
               keeplastbox,   
               ]{jacow}
\usepackage{pdfpages,multirow,ragged2e} %
\makeatletter%
	\ifboolexpr{bool{xetex}}
	 {\renewcommand{\Gin@extensions}{.pdf,%
	                    .png,.jpg,.bmp,.pict,.tif,.psd,.mac,.sga,.tga,.gif,%
	                    .eps,.ps,%
	                    }}{}
\makeatother

\ifboolexpr{bool{xetex} or bool{luatex}} 
 {}                                      
 {\usepackage[utf8]{inputenc}}           

\usepackage[USenglish]{babel}

\ifboolexpr{bool{jacowbiblatex}}%
 {%
  \addbibresource{TUPA095.bib}

 }{}
\begin{document}

\title{Techniques to seed the self-modulation instability of a long proton bunch in plasma}

\author{L. Verra\thanks{livio.verra@cern.ch}, G. Zevi Della Porta\textsuperscript{1}, E. Gschwendtner, CERN, Geneva, Switzerland\\
		M. Bergamaschi, P. Muggli, Max Planck Institute for Physics, Munich, Germany  \\
		\textsuperscript{1}also at Max Planck Institute for Physics, Munich, Germany}
	
\maketitle

\begin{abstract}
The Advanced Wakefield Experiment (AWAKE) at CERN relies on the seeded Self-Modulation (SM) of a long relativistic proton bunch in plasma to accelerate an externally injected MeV witness electron bunch to GeV energies.
During AWAKE Run 1 (2016-2018) and Run 2a (2021-2022), two seeding methods were investigated experimentally: relativistic ionization front seeding and electron bunch seeding.
In the first one, a short laser pulse copropagates within the proton bunch and ionizes the rubidium vapor, generating the plasma.
In the second, a short electron bunch propagates in plasma ahead of the proton bunch and drives the seed wakefields.
Both seeding methods will be further employed during AWAKE Run 2b (2023-2024) to study their effect on the SM evolution in the presence of a plasma density step.
In this contribution, we will show the main experimental results and discuss their impact for the future design of the experiment, in particular for Run 2c (starting in 2028), where the plasma will be split in two sections: one dedicated to SM of the proton bunch, and the other to the electron acceleration process.

\end{abstract}

Proton bunches produced routinely by synchrotrons have large energy per proton (GeVs to TeVs) and high charge per bunch (>10\,nC), resulting in a large amount of energy stored per bunch (>20\,kJ).
They can therefore drive wakefields~\cite{PWFA:CHEN} in plasma over long distance (10-$10^3\,$m), potentially leading to high energy gain (1-100s\,GeV)   by a witness electron ($e^-$) bunch in a single accelerating section, avoiding the complications of staging.
This was demonstrated with numerical simulations~\cite{ALLEN:NATURE} using a short $p^+$ bunch (root mean square bunch length $\sigma_z=100\,\mu$m).

\par However, $p^+$ bunches routinely produced, for example at CERN, are cm-long, much longer than the plasma electron wavelength in a plasma with electron density interesting for high-gradient acceleration.
Since conventional bunch compression~\cite{Assmann:1208433} and shaping~\cite{PATRIC:MASK} techniques are unpractical with relativistic protons, the transverse occurrence of the two-stream instability, the self-modulation (SM) instability~\cite{KUMAR:GROWTH, PUKHOV:GROWTH}, is used to form a train of short microbunches that can resonantly drive wakefields with amplitude to the GV/m level.

\par
\begin{figure*}[!h]
    \centering
    \includegraphics*[width=\textwidth]{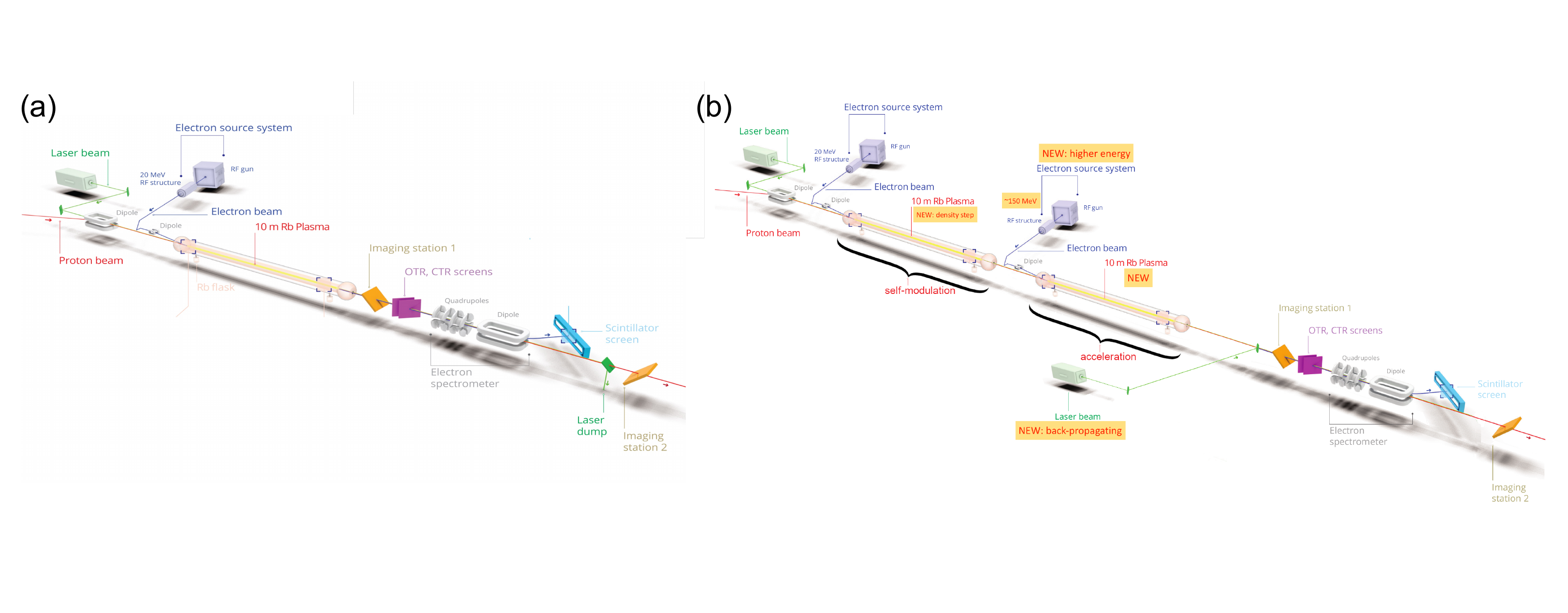}

    \caption{Experimental setup of AWAKE Run 1, 2a-b (a) and Run 2c (b).}
    \label{fig1}
\end{figure*}

 Large energy gain by a witness bunch requires a long plasma. 
In the Advanced Wakefield Experiment (AWAKE)~\cite{PATRIC:READINESS}, we use field-ionization of rubidium vapor by a $\sim10^{12}\,$W/cm$^2$ laser pulse to produce the plasma in a 10-m-long source (see Fig.~\ref{fig1}(a)). 
In previous experiments, we demonstrated that the 6-cm-long, 400\,GeV/c, 48\,nC $p^+$ bunch provided by the CERN Super Proton Synchrotron (SPS) self-modulates~\cite{KARL:PRL, MARLENE:PRL} and that electrons can be injected and accelerated to energies >2\,GeV~\cite{AW:NATURE}.
However, laser ionization does not scale favourably to very long length (20-$10^3\,$m), in particular because of energy depletion of the pulse by the ionization process. 
We therefore explore other plasma sources (e.g., discharge~\cite{TORRADO:DISCHARGE,} and helicon~\cite{HELICON:Buttenschön_2018}) that produce a preformed plasma.

\par Self-modulation is by nature an instability and must be seeded to reliably exploit it for reproducible particle acceleration.
Moreover, for the future design of the experiment (see Fig.~\ref{fig1}(b)), we foresee to use a first plasma (the "modulator") dedicated to the seeded SM of the $p^+$ bunch, and a second one (the "accelerator") dedicated to the acceleration of the witness bunch.
The two plasmas are separated by a $\sim 1$-m-long gap. 
This scheme is chosen to inject the 150\,MeV witness $e^-$ bunch on axis~\cite{LIVIO:EAAC} and after SM has developed and saturated in the first plasma to produce a microbunch train.
In addition, the first plasma will be equipped with a density step~\cite{PATRIC:EAAC} to avoid the decaying of the amplitude of the wakefields due to their dephasing with respect to the microbunch train~\cite{PUKHOV:GROWTH,SCH:GROWTH}.

\par In previous experiments we explored two seeding methods: relativistic ionization front (RIF)~\cite{FABIAN:PRL} and $e^-$ bunch seeding~\cite{LIVIO:PRL}.
In the following, we summarize the experimental results and discuss the advantages and disadvantages of each method, and their suitability for the future design of the experiment and for a high-energy accelerator for particle physics~\cite{AWAKE:sym14081680}.

\section{Relativistic Ionization Front Seeding}

\begin{figure}[!htb]
   \centering
   \includegraphics*[width=1\columnwidth]{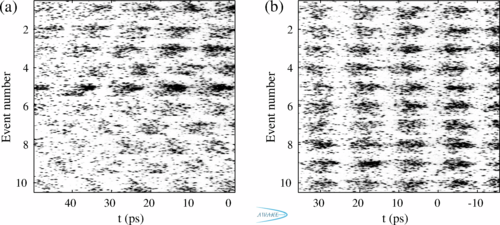}
   \caption{Ten consecutive time-resolved images of the self-modulated $p^+$ bunch in case of RIF propagating 600\,ps (a) and 350\,ps (b) ahead of the bunch center.
The bunch propagates from left to right.
$n_{pe}=0.94\times10^{14}\,$cm$^{-3}$.
Figure reproduced from~\cite{FABIAN:PRL}.}
   \label{fig2}
\end{figure}

Laser ionization makes RIF seeding possible: when the ionizing laser pulse copropagates within the $p^+$ bunch, the fast onset of the bunch-plasma interaction drives the initial seed wakefields, from which SM grows exponentially. 
Figure~\ref{fig2} shows two sets of consecutive time-resolved images of the $p^+$ bunch after propagation in plasma.
In Figure~\ref{fig2}(a), SM is not seeded: the timing of the microbunch train (traveling from left to right) is not reproducible from event to event.
In Figure~\ref{fig2}(b), SM is seeded: each microbunch appears at the same time along the bunch for all events.
Averaging the images of each set, one obtains a blurred image in the case of the instability, and a high-contrast images in the case of seeding (e.g., Fig.~\ref{fig3}(b)), depending on whether the underlying distribution is reproducible throughout the events or not.
The transition from the instability to the seeded regime occurs when the bunch density at the ionization front location is high enough to drive seed wakefields with amplitude $>4\,$MV/m~\cite{FABIAN:PRL}.
Thus, in the seeded case, there is a part of the bunch ahead of the RIF that keeps propagating as in vacuum.

\par This method has the advantage of the inherited alignment of the $p^+$ bunch with the plasma column, which makes it simpler in terms of operation.
The seed wakefields act symmetrically on the bunch, and therefore the undesired asymmetric counterpart of SM, the hosing instability~\cite{WHITTUM:PhysRevLett.67.991}, is suppressed~\cite{VIEIRA:PhysRevLett.112.205001}.

\par However, in the future setup of the experiment (Fig.~\ref{fig1}(b)), the front of the bunch, left unmodulated after the first plasma, enters the second (preformed) plasma diverging and with large transverse size.
In case the front of the bunch self-modulates in the second plasma, the wakefields that it drives may disrupt the structure of the (seeded) self-modulated back and spoil the acceleration process.

\par Recent experimental results~\cite{LIVIO:WIDE} indicate that, over 10\,m of propagation in plasma, a bunch with transverse size and divergence comparable to that of the bunch front entering the second plasma does not self-modulate ahead of the transition point along the bunch. 
Therefore, RIF seeding remains a viable option for AWAKE Run 2c and future accelerator schemes.

\section{Electron Bunch Seeding}

\begin{figure}[!htb]
   \centering
   \includegraphics*[width=1\columnwidth]{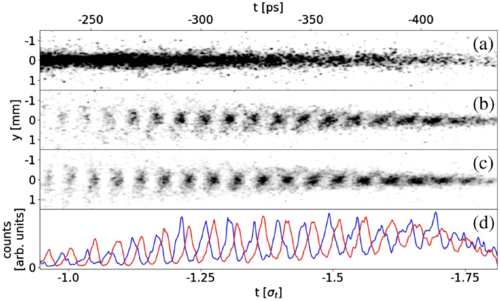}
   \caption{Time-resolved images of the $p^+$ bunch at a screen emitting optical transition radiation and positioned 3.5\,m downstream of the plasma exit. 
   Each image is the average of 10 single-event images. 
Bunch center at t=0\,ps, the bunch
travels from left to right. 
(a) No plasma (incoming bunch). 
(b) Plasma ($n_{pe}=1.02 \times 10^{14}\,$cm$^{-3}$) and seed $e^-$ bunch.
(c) Same as (b) but $e^-$ bunch delayed by 6.7 ps
All images have the same color scale.
(d) On-axis time profiles of (b) (blue line) and (c) (red line) obtained by summing counts over -$0.217\leq y \leq 0.217\,$mm.
Figure reproduced from~\cite{LIVIO:PRL}.}
   \label{fig3}
\end{figure}

The initial seed wakefields can also be generated by a charged particle bunch propagating ahead of the $p^+$ bunch.
In AWAKE, we used the short 19\,MeV $e^-$ bunch previously used for acceleration experiments to demonstrate this experimentally.
In this case, the ionizing laser pulse travels ahead of both electron and proton bunches and preforms the plasma.
Figure~\ref{fig3} shows averaged time-resolved images of the front of the $p^+$ bunch propagating in vacuum (a), and in plasma with the seed $e^-$ bunch (b,c).
The high contrast of averaged time-resolved images with plasma confirms that the timing of the microbunch train (i.e. of the wakefields) is reproducible from event to event with electron bunch seeding.
We also demonstrated that the timing of the modulation is tied to the relative timing between the seed $e^-$ and $p^+$ bunch by delaying the seed by $6.7\,$ps (close to half a plasma electron period, for the density used).
This results in a shift in time of the microbunch train by the same amount (Fig.~(c)), that is clearly visible from the on-axis profiles of the two images (Fig.~(d), blue line: profile of (b), red line: profile of (c)).

\par This method has the advantage of applying the seed wakefields on the entire $p^+$ bunch, which would enter the second plasma fully self-modulated with reproducible timing from event to event.
In fact, Figs.~\ref{fig3}(b,c) show that the bunch  self-modulates from the very front of the bunch.
Moreover, the amplitude of the seed wakefields and the growth rate of the instability can be varied independently by varying the parameters of the $e^-$ or $p^+$ bunch, respectively, allowing for additional control on the development of the process~\cite{LIVIO:PRL}.

\par The disadvantage of this method is that it requires aligning the $e^-$ bunch trajectory, both in position and angle, with the plasma column and with the $p^+$ bunch trajectory. 
Recent experimental results~\cite{TATIANA:HOSING} show that, when the transverse seed wakefields act asymmetrically on the $p^+$ bunch, the hosing instability, which is detrimental for the acceleration process, arises.
Electron bunch seeding is also less practical than RIF seeding because it requires an additional electron source and beamline to provide the seed bunch.
\section{AWAKE Run 2 Experimental Program}

The experimental program of AWAKE Run 2 is divided in phases, each dedicated to particular physics milestones. 
The final goal is to generate high-quality and high-energy $e^-$ bunches suitable for particle physics experiments. 
During Run 2a (2021-2022), we successfully demonstrated the $e^-$ bunch seeding~\cite{LIVIO:PRL}, studied the development of hosing instability~\cite{TATIANA:HOSING} and we further tested the RIF seeding scheme~\cite{LIVIO:WIDE}. 
Run 2b (2023-2024) will investigate the evolution of SM in the presence of a step in the plasma electron density.
Numerical simulations~\cite{KOSTANTIN:STEP} show that a sudden increase in plasma electron density limits the dephasing of wakefields with respect to the microbunch train and hence the decaying of the amplitude of the wakefields. 
This is important to maintain a high accelerating gradient for long distances in plasma.
The evolution of the wakefields along the plasma will be studied by measuring the plasma recombination light~\cite{PLASMA_LIGHT:Oz2004ts,PATRIC:AAC22} and the energy spectrum of accelerated electrons.
Both seeding methods will be further explored in the presence of the density step.

\par Run 2c (to start in 2028) requires major modifications of the existing facility, with the installation of 
a second plasma source, as well as an electron source and beamline providing the 150\,MeV witness $e^-$ bunch.
The experiment will focus on acceleration of $e^-$ bunches in the second plasma with control of the bunch quality.
Both seeding methods will be tested again, in order to determine the final design of a plasma wakefield accelerator for applications, based on the self-modulation scheme.
The following Run 2d will employ scalable plasma sources to reach even higher energy gains, on the path towards particle physics applications.



%
%
\ifboolexpr{bool{jacowbiblatex}}%
	{\printbibliography}%
	{%
	

} 

\end{document}